
\magnification=\magstep1
\baselineskip=14pt
\def\nms{\mathsurround=0pt}
\def\overapprox#1#2{\lower2pt\vbox{\baselineskip0pt\lineskip - 1pt
    \ialign{$\nms#1\hfil##\hfil$\crcr#2\crcr\approx\crcr}}}
\def\gtsim{\mathrel{\mathpalette\oversim>}} 
\def\ltsim{\mathrel{\mathpalette\oversim<}} 
\def\oversim#1#2{\lower2pt\vbox{\baselineskip0pt\lineskip 1pt
    \ialign{$\nms#1\hfil##\hfil$\crcr#2\crcr\sim\crcr}}}

\magnification=\magstep1
\baselineskip=14pt

\def \nc1 {\nu _c^{-1}}
\centerline{\bf CHAOS IN A HOMOGENEOUS MODEL FOR EARTHQUAKES}
\vskip 3cm
\centerline {Maria de Sousa Vieira}
\bigskip
\centerline{\sl Hochstleistungsrechenzentrum Supercomputing Center,
Kernforschungsanlage}
\centerline{\sl D-52425 J\"ulich }
\centerline{\sl Germany}
\vskip 3cm
\centerline {ABSTRACT}
\bigskip
{\sl We investigate the nonlinear properties  of
a system introduced
by Burridge and Knopoff to model the dynamics of earthquakes. We find
that a two-block system in a completely homogeneous configuration presents
a complex behavior characterized by the presence of
periodic, quasiperiodic and chaotic orbits.
We have found routes to
chaos via two types of intermittencies and period doubling bifurcations.
The sensitivity of the evolution to different initial conditions
is quantified by calculating the largest Liapunov exponent.
The dynamics of the model is governed by nondifferentiable flows. }
\bigskip
\medskip
PACS numbers: 05.45+b, 91.45.Dh, 91.60Ba.
\vfill\eject
Earthquakes are catastrophic events that happen when tectonic plates
move with respect each other. The movement of the plates occurs in an
intermittent way. A short period of slip follows a long period of rest, and
in its turn is followed again by a short slipping period. This kind of dynamics
is
called stick slip motion.

In 1967 Burridge and Knopoff [1] introduced two simple models
that mimic the dynamics of earthquakes. The systems consist
of blocks connected by springs. The set is pulled with constant
velocity on a surface with friction.
In one of these models only the first block is coupled to  the
driving mechanism. It has been called ``the train model"[2].
In the
other model, all the blocks are connected to the driving element[1,3,4].
We
call it here the BK model. It is observed that in a completely
homogeneous configuration of their component elements, both models
present stick slip dynamics, and the
distribution of the slipping events follows a power-law,
in qualitative agreement with the distribution observed
in real earthquakes, that is, the Gutenberg-Richter law[5].
However, there are fundamental differences between the train model
and the BK model. In the first one the power-law distributions
are limited only by the size of the system, that is, it is observed
the existence of self-organized criticality[6]. In the BK model
this does not occur.
The power-laws have a limited extent[3,4] and they seem to result from finite
size effects[7].

The studies on the Burridge-Knopoff models have been mostly
concentrated to big chains, since the primary aim of these
studies have been the comparison of the statistical distributions
of the slipping events with the distribution of real earthquakes.
In recent publications, the dynamics of a small chain of the
BK model was
investigated. Huang and Turcotte have shown that if an
asymmetry is introduced in the friction forces,
then a system of two blocks can present chaotic behavior[8,9].
More specifically, they considered the case in which the friction force
in one block is different from the friction force in the other block.
They have compared their results with the motion of tectonic
plates in California and in Japan
and observed that they seem to have the same kind of chaotic dynamics
observed in the two-block system of the BK model[9]. To the
best of our knowledge, chaos has not been seen in the two-block
system of the homogeneous BK model
if no asymmetry is introduced in it.
A one-block system, both in the BK and train models, does not present chaos. In
this case only a periodic behavior is observed.

The motivation for studying these mechanical models for earthquakes
is in part to investigate one important open question in
seismology, that is,
are the temporal and spatial complexity of earthquakes
generated by the nonlinearities of the equations that govern these
phenomena? or  are they due to geometrical and built-in heterogeneities?
Another motivation results from the fact that these models
are described by flows that are not infinitely differentiable,
like most of the systems studied in chaos theory.
Little is known about the class of systems we consider here,
both from the
physical and mathematical points of view.

The aim of this paper is to investigate the dynamics of a small
system in the other model introduced by Burridge and Knopoff,
i.e., the train model. In this case the two blocks would
also represent two coupled tectonic faults. We find that {\sl without any kind
of
asymmetry, chaotic behavior is ubiquitous in a  two-block system}.
We observe a rich bifurcation diagram characterized by periodic, quasiperiodic
and chaotic trajectories. We find entrances into chaos
via intermittencies of types I and II and period doubling bifurcations.
The phenomenon of crisis is also seen.
To quantitatively characterize the divergence (or convergence) of
nearby trajectories
we  calculate the most important (namely, the largest) Liapunov exponent of the
system.

The model we study is shown schematically in Fig. 1. It is the
train model where the number of blocks is two.
Each block of the system has mass $m$ and the springs in the model are
characterized by an elastic constant
$k$. The first block is pulled with constant velocity
$v $ and the friction force $F$ between the blocks  and the
surface is a function of the instantaneous velocity of the block
with reference
to a characteristic velocity $v_c$.

The equations of motion for the first and second blocks are given respectively
by
$$m\ddot X_1=k(X_{2}-2X_1+vt)-F(\dot X_1/v_c), \eqno (1a) $$
$$m\ddot X_2=k(X_{1}-X_2)-F(\dot X_2/v_c), \eqno (1b) $$
where $X_j$ denotes
the displacement of the  block measured
with respect to the position where the
sum of the elastic
forces in it is zero. These equations are applicable only when the
respective block is moving and the sum of the elastic forces in
the block is larger than the maximum force of static friction.
If this condition is not met we simply have $\dot X_j=0$.
If we write the friction force as
$F(\dot X_j/v_c)= F_{\circ}\Phi(\dot X_j/v_c) $
where $\Phi (0)=1$
and introduce the variables
$\tau =\omega _p t,\ \ \omega _p^2=k/m,\ \ U_j=kX_j/F_{\circ}$,
Eqs. (1) can be written in the following dimensionless form[2]
$$\ddot U_1=U_{2} -2U_1+\nu \tau-\Phi(\dot U_1/\nu_c), \eqno (2b)$$
$$\ddot U_2=U_{1} -U_2-\Phi(\dot U_2/\nu_c), \eqno (2b)$$
with $\nu=v/V_{\circ}$,
$\nu_c=v_c/V_{\circ}$, and $V_{\circ}=F_{\circ}/\sqrt{km}$.
Dots now denote differentiation with respect to $\tau $. In a system
of a single block the quantity $F_{\circ}/\omega _p$ is the
maximum displacement of the pulling spring before the block
starts to move; in the absence of dynamical friction
$2\pi/\omega _p$  and $V_{\circ}$ are respectively
a characteristic period of oscillation of the block and the maximum velocity it
attains. We find that this system
is completely described by  two dimensionless parameters, $\nu $ and $\nu _c$.
The system is four-dimensional, since its evolution is completely
specified by giving the initial positions and velocities of the two
blocks.
We use here the velocity weakening friction force given by[3]
$$\Phi (\dot U/\nu _c)= {{sign(\dot U)}\over{1+|\dot U|/\nu _c}}, \eqno(3)$$
which is a simple nonlinear function. The friction force is
the only nonlinear element in this model.

A possible solution for the motion of the system is when both blocks
move with constant velocity, equal to the pulling velocity $\nu $.
The solutions for the positions of the blocks found from Eq. (2) are in this
case
$$U_1^e=-2/(1+\nu/\nu _c)+\nu \tau, \eqno (4a)$$
$$U_2^e=-3/(1+\nu/\nu _c)+\nu \tau, \eqno (4b)$$
where the superscript $e$
denotes equilibrium position.
The stability of this solution can by investigated by calculating
the eigenvalues of the Jacobian matrix of a
transformed system in which the equations are first order ODE's. This
can be done by introducing
two extra variables, $V_1\equiv \dot U_1$ and $V_2\equiv \dot U_2$, which allow
us to write
the equations of motion as
$$\eqalign {\dot U_1&=V_1\cr
\dot V_1&=U_{2} -2U_1+\nu \tau-\Phi(V_1/\nu_c) \cr
\dot U_2&=V_2\cr
\dot V_2&=U_{1} -U_2-\Phi(V_2/\nu_c) \cr}. \eqno (5)$$
The Jacobian matrix of this system when $V_1=V_2=\nu$ has the
eigenvalues given by
$$\lambda _i= {{A\pm  \sqrt{B_{\pm}} }\over {2}}, \eqno (6)$$
where $i=1,...,4$,
$A=\nu _c/(\nu _c+\nu )^2$
 and $B_{\pm }=-6\pm 2\sqrt {5} + A^2$. For the region where we
concentrate our attention ($0 < \nu \ltsim1$ and $ 0< \nc1 \ltsim 1$)
we find that $B_+$ and $B_-$ are always smaller than zero.
This gives two pairs of complex eigenvalues with
$$|\lambda _1|=|\lambda _2|=\sqrt{{{3 -  \sqrt {5}}\over {2}}}\eqno (7a)$$
$$|\lambda _3|=|\lambda _4|=\sqrt{{{3 +  \sqrt {5}}\over {2}}}\eqno (7b)$$
Since $|\lambda _{3,4}| >1$, this implies that
the equilibrium point is unstable.  In fact, it is a saddle point
with two stable directions and two unstable directions.
Therefore, if this solution is
perturbed, the motion settles
into another attractor.

For a random initial condition,
we observe that the stick slip dynamics is the typical situation for
the dynamical evolution of our system.
Depending on the used parameters,
we can find quasi-periodic, periodic or chaotic behavior.
Examples  for these
types of motion are shown in Fig. 2.
In all the numerical simulations shown here
we start the system
with  the blocks
at rest. The initial position for each block is
$U_j=0$, i.e., with the sum of the elastic forces in the
block equal to zero. Thus, no randomness is introduced, {\sl  not even in
the initial positions or velocities of the blocks}.
A transient time of $\tau =3000$ is discareded in all simulations.
Figs. 2(a) and 2(b) show the system evolution in phase-space for
the first and second blocks, respectively, in a case
of quasi-periodic motion where $\nc1 =0.03$.
In Figs. 2(c) and 2(d) we show the dynamical evolution of the blocks
when $\nc1 =0.05$, which gives a periodic motion.
Note that for this case, only the first
block sticks to the surface.
Finally, Figs. 2(e) and 2(f) show the evolution when
$\nc1 =0.39$, which gives a chaotic orbit.
In all cases of Fig. 2 we have taken the pulling velocity  $\nu =0.1$.
It is clear that the characterization of the motion as chaotic or
nonchaotic can be done only with a more careful study, by
investigating the Liapunov exponents. This has been done and
will be discussed in the next paragraphs.

Note that we have plotted in the $x$-axis the position of the block
with respect to its (unstable) equilibrium position $U^e_j$, since
the evolution of each block occurs around its respective equilibrium point.
As seen in the figures, there is a discontinuity in $\ddot U_j$ when
the block sticks to the surface. So, this flow is not
infinitely differentiable, like most
of the flows considered in the literature of
dynamical systems.

By varying the degree of nonlinearity, i.e., varying $\nu _c$,  and fixing the
pulling velocity
we investigate how the motion
changes its character from a periodic to a nonperiodic case. This
is usually done by plotting  bifurcation diagrams.
For our  case, a
Poincar\'e section [10]
will lower the dimensionality
of the system from four to three. Even in this situation, the visualization of
the bifurcation diagrams is still complicated.
We have verified that for our
system the motion of the center-of-mass gives
a good description of the dynamics of the model.
Thus, we study  bifurcation diagram by investigating the
evolution of the center-of-mass in a determined Poincar\'e section.
However, we do not expect that this approach would always be good
for a system with very large dimensionality.

The coordinate of the center-of-mass with respect
to the equilibrium position, determined by Eq. 4, is
$W=(U_1-U^e_1+U_2-U^e_2)/2$.
We take Poincar\'e section of $\dot W$ at
$W=0$. In this way, we reduced the dynamics to a study of  a one-dimensional
variable.
We show in Fig. 3 a bifurcation diagram for $\nu =0.1$.
On the $x$-axis we
have $\nc1 $ and on the $y$-axis we plot $\dot W $ at
$W=0$. In the diagram we can see windows of
periodic motion, and regions where the motion
is nonperiodic.
We find regions with period doubling bifurcation route
to chaos, but also, as described later, other
routes to chaos are also present in our system.
We  have studied bifurcation diagrams for other pulling
velocities, and verified that their topologies can be different
from the case we show here. These studies will be published elsewhere [11].

The nature of the nonperiodic
motion can be analyzed by studying the sensitivity of the dynamics to
different nearby initial conditions. If this sensitivity
is found, we have by definition a chaotic motion.
Otherwise, the dynamics will be quasiperiodic.
It is clear that in a periodic motion two nearby initial
conditions will converge to the same orbit (if they
are in the same basin of attraction).
In Fig. 4 we show the separation of two nearby orbits of
the chaotic trajectory shown in Fig. 2(e). After the transient
period dies out, we add a small perturbation of $10^4$ to the positions
and velocities of the blocks. At the instant of perturbation
we have $U_1-U_1^e=-1.72$ and $\dot U_1=0.09$.
We can see from the figure that at this point the two orbits
practically coincide with each other. After a quite short
time of integration ($\tau =60$) presented in the figure, the two orbits have
diverged
considerably from each other.

The quantity generally used to
characterize the divergence (or convergence) of
nearby trajectories is the Liapunov exponent.
If the largest Liapunov exponent $\lambda _m$ of the system is positive, then
we
have a chaotic motion. The calculation of $\lambda _m$ can be
done in the following way:
After the transient dies out,
we give small perturbations to the orbit
and
verify how much it separates after a small
time interval from the original unperturbed orbit.
The perturbations are given in the direction of
the largest orbit separation.
The logarithm of the average orbit separation along the trajectory, divided by
the
time interval gives $\lambda _m$.
This standard method of calculating the largest
Liapunov exponent of a system was introduced in [12].

In Fig. 5 we plot
$\lambda _m$ for the bifurcation diagram shown in Fig. 3.
When the nonlinearity is small,
that is, $\nc1 $ is small, the motion is found
to be periodic and quasiperiodic,  as
expected. For a flow this implies $\lambda_{m}=0$,
and the distinction of a  motion with a large period and a quasi-periodic orbit
can be done by calculating the second largest Liapunov exponent.
However, here we concentrate our attention only to $\lambda _m$.
The big period-three window
loses stability at $\nc1 =0.146$ and after this the
system enters into chaos.
For $\nc1 \gtsim 0.146$ the
orbit stays for a long period of time close to the old period-three
trajectory, and then makes chaotic incursions into other regions.
That is, here we see what is called intermittency road to chaos. The
intermittency is of type II, according
to the classification given by Pomeau and Manneville [13].
It seems difficult to calculate explicitly the eigenvalues
of the map at the Poincar\'e sections  and in this
way verify how they cross the unit circle, characterizing
the intermittency.
However, by studying the
third Poincar\'e return map,
we see the typical structure of this
type of intermittency. Details about this will be reported in a
future publication [11].
After the first entrance into chaos,  there exist several
windows of periodic motion,
giving $\lambda _m=0$, as Fig. 5 shows.

Other rich phenomena are also observed in the bifurcation
diagrams. Period doubling cascades are born
in tangent bifurcations, and present the well known
Feigenbaum's exponents [14]. A big periodic
window is born in a tangent
bifurcation at $\nc1 =0.270$. For $\nc1 \ltsim 0.270$ there is an
entrance into chaos via
intermittency of type I [13].
Other  type I
intermittencies are seen at $\nc1 =0.498$ and $\nc1 =0.640$. At $\nc1 \approx
0.540$
we see a crisis-induced intermittency [15].
That is, for $\nc1 \gtsim 0.540$ the attractor suddenly
widens. It spends long stretches of time in the region
to which it was confined for $\nc1 \ltsim 0.540$. At the end
of these  long stretches the orbit bursts out of the
old region and bounces around chaotically in the new
enlarged region made available to it by the crisis.
Another crisis induced intermittency is seen at
$\nc1 \approx 0.683$.
For larger values of $\nc1 $
practically only chaotic behavior is seen.

In conclusion, we have studied the dynamics of a two-block system in
a one-dimensional
model for earthquake. We have found a rich dynamics with
periodic, quasiperiodic and chaotic motion.  The system presents
several
routes to chaos, as intermittencies of type I and II and
period doubling bifurcations.
A question we are currently investigating is whether or not the system
becomes more chaotic as the number of block increases. We have
seen that this model with large number of blocks presents self-organized
criticality[2], and it
has been claimed that self-organized critical systems are not
chaotic[16]. However, this contradicts the common result that
a system becomes more chaotic as its dimensionality increases. If
chaos is a general solution for this model
and if this system
gives a reasonable description of the irregular dynamics observed in
real earthquakes, this could mean that in practice earthquakes
are  predictable only on a short time basis, due to the sensitivity
of the evolution
to different initial conditions.
\medskip
\bigskip
\noindent{ACKNOWLEDGMENTS}
\medskip
\medskip
I thank the Alexander von Humboldt fundation for financial support
and H. J. Herrmann for the warm hospitality at the
Hochstleistungsrechenzentrum.
\bigskip
\medskip
\noindent {REFERENCES}
\bigskip
\item {1.} R. Burridge and L. Knopoff, {\sl Bull. Seismol. Soc. Am.} {\bf 57},
341 (1967).

\item {2.} M. de Sousa Vieira, {\sl Phys. Rev. A} {\bf 46}, 6288 (1992).

\item {3.} J. M. Carlson and J. S. Langer, {\sl Phys. Rev. Lett.} {\bf 62},
2632 (1989); {\sl Phys. Rev. A} {\bf 40}, 6470 (1989).

\item {4.} G. Vasconcelos, M. de Sousa Vieira and S. R. Nagel, {\sl Physica
A} {\bf 191}, 69 (1992); M. de Sousa Vieira, G. Vasconcelos and S. R. Nagel,
{\sl Phys. Rev. E} {\bf 47}, R2221 (1993).

\item {5.} B. Gutenberg and C. F. Richter, {\sl Bull. Seismol. Soc. Am.}
{\bf32},
163 (1942); {\sl Seismicity of the Earth and Associated Phenomena} (
Princeton University Press, Princeton, N. J., 1954).

\item {6.} P. Bak, C. Tang, and K. Wiesenfeld, {\sl Phys. Rev. Lett.}
{\bf 59}, 381 (1987); {\sl Phys. Rev. A} {\bf 38}, 364 (1988).

\item {7.} J. Schmittbuhl, J-P. Vilotte and S. Roux,
{\sl Europhys. Lett.} {\bf 21}, 375 (1993).

\item {8.} J. Huang and D. Turcotte, {\sl Geophys. Res. Lett.} {\bf 17},
223 (1990).

\item {9.} J. Huang and D. Turcotte, {\sl Nature} {\bf 348}, 234 (1990).

\item {10.} See for example,
A. J. Lichtenberg, and M. A.  Lieberman, M. A.,
``{\sl Regular and Stochastic Dynamics}" (Springer, New York, 1992).

\item {11.} G. Benettin, L. Galgani, and J. M. Strelcyn, {\sl Phys.
Rev. A} {\bf 14} 2338 (1976).

\item {12.} Maria de Sousa Vieira, {\sl in preparation}.

\item {13.} Y. Pomeau and P. Manneville, {\sl Commun. Math. Phys.}
{\bf 74}, 189 (1980).

\item {14.} M. J. Feigenbaum, {\sl J. Stat. Phys.} {\bf 19}, 25 (1978).

\item {15.} C. Grebogi, E. Ott, F. Romeiras and J. A. Yorke, {\sl Phys.
Rev. A} {\bf 36} 5365 (1987).

\item {16.} P. Bak, K. Chen in {\sl ``Nonlinear Structure in Physical
Systems - Pattern Formation, Chaos and Waves",} edited by
L. Lam and H. C. Morris (Springer, 1990).
\vfill\eject
\noindent{FIGURE CAPTIONS}
\medskip
\medskip
\item {Fig. 1.}  System studied, which  consists of a chain of two blocks
connected by linear springs. The blocks are on a flat surface and the
first one is pulled with constant velocity. Between the surface and
the blocks there is velocity weakening friction force.

\item {Fig. 2.} Orbits in phase-space for each block when $\nu =0.1$.
In (a) and (b) we have $\nc1 =0.03$, which results in a quasi-period  orbit.
For (c) and (d) we have $\nc1 =0.05$, which gives in a periodic trajectory.
Finally, in  (e) and (f) we have $\nc1 =0.39$, giving a chaotic motion.
The vertical dotted lines just indicate the equilibrium position for the
respective block.

\item {Fig. 3.} Bifurcation diagrams of the velocity of center-of-mass $\dot W$
on
the surface of section $W=0$ as a function of $\nc1 $ with $\nu =0.1$.

\item {Fig. 4.} Separation of two nearby orbits for the chaotic
trajectory shown in Fig. 2(e) where $\nu =0.1$ and $\nc1 =0.39$. A
perturbation of $10^4$ is given to the positions and velocities
of the blocks when
$U_1-U^e_1=-1.72$ and $\dot U_1=0.09$. The integration time for this figure is
$\tau =60$,
after discarding the transient period.

\item {Fig. 5.} The largest Liapunov exponent corresponding to the
bifurcation diagram shown in Fig. 3, where $\nu =0.1$. The calculation of
$\lambda _m$
is done for an integration time $\tau =15000$, with time steps of
$\Delta \tau =0.01$ and perturbations to the position and
velocities of the blocks equal to $10^{-5}$.

\end